| Title | **Cold atmospheric plasma decontaminates Arabidopsis thaliana seeds and remodels seedling fungal microbiota** |
|---|---|

| **Authors** | Léna Taras[1,2,3], Nicole Chaumont[1,2], Caroline Kunz[4,5], Thierry Dufour[3,†] and Christophe Bailly[1,2,*,†] |
|---|---|
| **Affiliations** | [1]Development, Adaptation and Ageing (Dev2A), Sorbonne Université, CNRS, Inserm, F-75005 Paris, France<br>[2]Institut de Biologie Paris-Seine (IBPS), Sorbonne Université, CNRS, Inserm, F-75005 Paris, France<br>[3]Laboratoire de Physique des Plasmas (LPP), Sorbonne Université, CNRS, École Polytechnique, F-75005 Paris, France<br>[4]Unité Molécules de Communication et Adaptation des Micro-Organismes, Muséum National d'Histoire Naturelle, UMR 7245, F-75005 Paris, France<br>[5]Faculté des Sciences et Ingénierie, Sorbonne Université, UFR 927, F-75005 Paris, France |
| **Correspondence** | E-mail: thierry.dufour@sorbonne-universite.fr |

| **Abstract** | Seed-associated microorganisms influence seed quality, seedling establishment, and plant health and also constitute a major source of seed-borne pathogens. Cold atmospheric plasma (CAP) has emerged as a promising alternative to chemical seed treatments because of its antimicrobial activity, although its effects on fungal communities associated with developing seedlings remain poorly understood. Here, Arabidopsis thaliana seeds from two ecotypes (Columbia and Landsberg erecta) were exposed to CAP for 5 or 15 min. Seed decontamination efficiency was assessed by culturing on malt extract agar and nephelometric analyses, while fungal communities associated with seedlings derived from treated and untreated seeds were characterized by ITS1 amplicon sequencing. CAP efficiently reduced fungal contamination without affecting seed germination. CAP altered the composition of fungal communities associated with developing seedlings, but the magnitude of these changes depended on seed batch and ecotype. Dominant taxa markedly declined after treatment, whereas several low-abundance taxa increased in relative abundance. These findings demonstrate that CAP is an effective pesticide-free technology for seed decontamination and can also reshape fungal communities associated with developing seedlings, highlighting broader ecological consequences of plasma-based seed treatments. |
|---|---|

# I. Introduction

Plant diseases represent a major constraint to agricultural productivity and are responsible for substantial crop losses worldwide. Their control relies heavily on the use of chemical pesticides, fungicides, and antimicrobial compounds. Over the last decades, the global consumption of these products has increased considerably, reflecting the growing pressure exerted by plant pathogens on agricultural systems [1]. Although effective, these conventional approaches raise important environmental concerns. Repeated applications of phytosanitary products can alter soil biological quality, reduce microbial diversity, promote the emergence of resistant microorganisms, and contribute to environmental contamination of soils and water resources [2], [3]. Consequently, the development of sustainable alternatives for plant disease management has become a major challenge for modern agriculture.

Seeds are the main propagules of most crop species and can act as vectors of both beneficial and detrimental microorganisms. Seeds harbor diverse microbial communities collectively referred to as the seed microbiota [4]. These communities include endophytic microorganisms residing within seed tissues and epiphytic microorganisms colonizing external seed surfaces [5]. The composition of the seed microbiota results from complex interactions between host genotype and environmental factors and contributes to the establishment of microbial communities throughout plant development [6]. Beyond their role in plant colonization, seed-associated microorganisms can directly influence seed quality, germination, seedling establishment, and plant health. The effects of plant-associated microbial communities on their host are highly context dependent. Depending on environmental conditions and community composition, microorganisms may exert beneficial, neutral, or detrimental effects on plant performance [7]. Numerous studies have highlighted the beneficial functions of plant-associated microbes, including growth promotion, nutrient acquisition, stress tolerance, biological control of pathogens, and the production of bioactive compounds [7], [8]. These properties have stimulated growing interest in the exploitation of microbiota-based approaches for sustainable agriculture and ecological restoration [1]. However, environmental conditions can profoundly modify microbial behavior and alter host–microbe interactions. For example, Hiruma et al. [10] demonstrated that Colletotrichum tofieldiae promotes Arabidopsis growth under phosphate-deficient conditions, whereas Finkel et al. [11] showed that phosphate limitation can shift Burkholderia species from a commensal to a pathogenic lifestyle. Such observations illustrate how environmental disturbances may disrupt microbial equilibrium and reveal pathogenic traits in microorganisms that remain harmless under favorable conditions.

Among the environmental factors affecting seed-associated microbial communities, post-harvest storage conditions are particularly important. Variations in storage duration, temperature, or moisture content can modify microbial abundance and community structure, influencing both seed quality and health [12]. Under unfavorable storage conditions, microorganisms already present on seeds may proliferate, leading to seed deterioration, reduced germination capacity, and increased risks of transmitting pathogens to the next plant generation [12]. Consequently, seed treatments capable of reducing microbial contamination while preserving seed quality represent an important objective for sustainable crop production.

Among the alternative technologies currently being explored, cold atmospheric plasma (CAP) has emerged as a promising approach for seed decontamination. Cold plasma is a partially ionized gas composed of highly energetic electrons, ions, reactive oxygen and nitrogen species, neutral particles, and photons. Unlike thermal plasmas, CAP operates close to room temperature because only the electron population reaches high energy levels, while the bulk gas remains near ambient temperature [13]. This unique property makes CAP particularly suitable for biological applications. Owing to the generation of reactive species, CAP exhibits strong antimicrobial activity and has been widely investigated for medical and industrial disinfection applications [14], [15]. In agriculture, numerous studies have demonstrated its ability to inactivate seed-borne pathogens and reduce microbial contamination without the use of chemical pesticides [16], [17]. These characteristics make CAP an attractive candidate for the development of environmentally friendly seed treatment technologies. Despite increasing interest in plasma-based seed treatments, their effects on the structure of seed-associated microbial communities remain poorly understood. Previous studies in Arabidopsis and sunflower have suggested that plasma treatment may modify bacterial communities associated with seedlings derived from treated seeds [18], [19]. Effects on seed-associated fungal communities have received less attention. In common and Tartary buckwheat, CAP treatment reduced the frequency and diversity of seed-borne fungi, with fungal taxa showing differential sensitivity to plasma exposure [20]. However, how CAP treatment of seeds affects the composition of fungal communities subsequently associated with developing seedlings remains poorly characterized.

We hypothesized that CAP treatment would efficiently reduce the microbial load associated with seeds and induce measurable changes in the composition of fungal communities associated with emerging seedlings. To test this hypothesis, we investigated the effects of CAP treatment on seeds of Arabidopsis thaliana ecotypes Columbia and Landsberg erecta. We first assessed the efficiency of CAP-mediated seed decontamination by monitoring microbial growth on culture medium, complemented by nephelometric analysis of fungal growth. We then used ITS1 amplicon sequencing to determine whether CAP treatment altered the composition and diversity of fungal communities associated with seedlings derived from treated seeds. The use of two ecotypes and two independent seed batches allowed us to assess the consistency of these responses across seed lots differing in their initial levels of microbial contamination.

# 2. Results

## 2.1. Effects of cold plasma treatment on seed contamination

### 2.1.1. CAP treatment reduces microbial growth from seeds on MEA medium

We first assessed the effect of CAP exposure on seed-associated microbial contamination by monitoring visible microbial growth from untreated and CAP-treated seeds on malt extract agar (MEA) medium, using two independent seed batches from the Col and Ler ecotypes. All seeds batches germinated to 100% within 3 days at 15 °C and CAP treatment did not markedly modify final germination percentage nor germination speed (**Supplemental Figure S1**). Without any treatment, seed contamination levels, evaluated after 6 days on MEA medium, ranged from 20 to 100% in Col seeds (**Figure 1A, 1B**) and from 80 to 100% in Ler seeds (**Figure 1C, 1D**), depending on the seed batch.

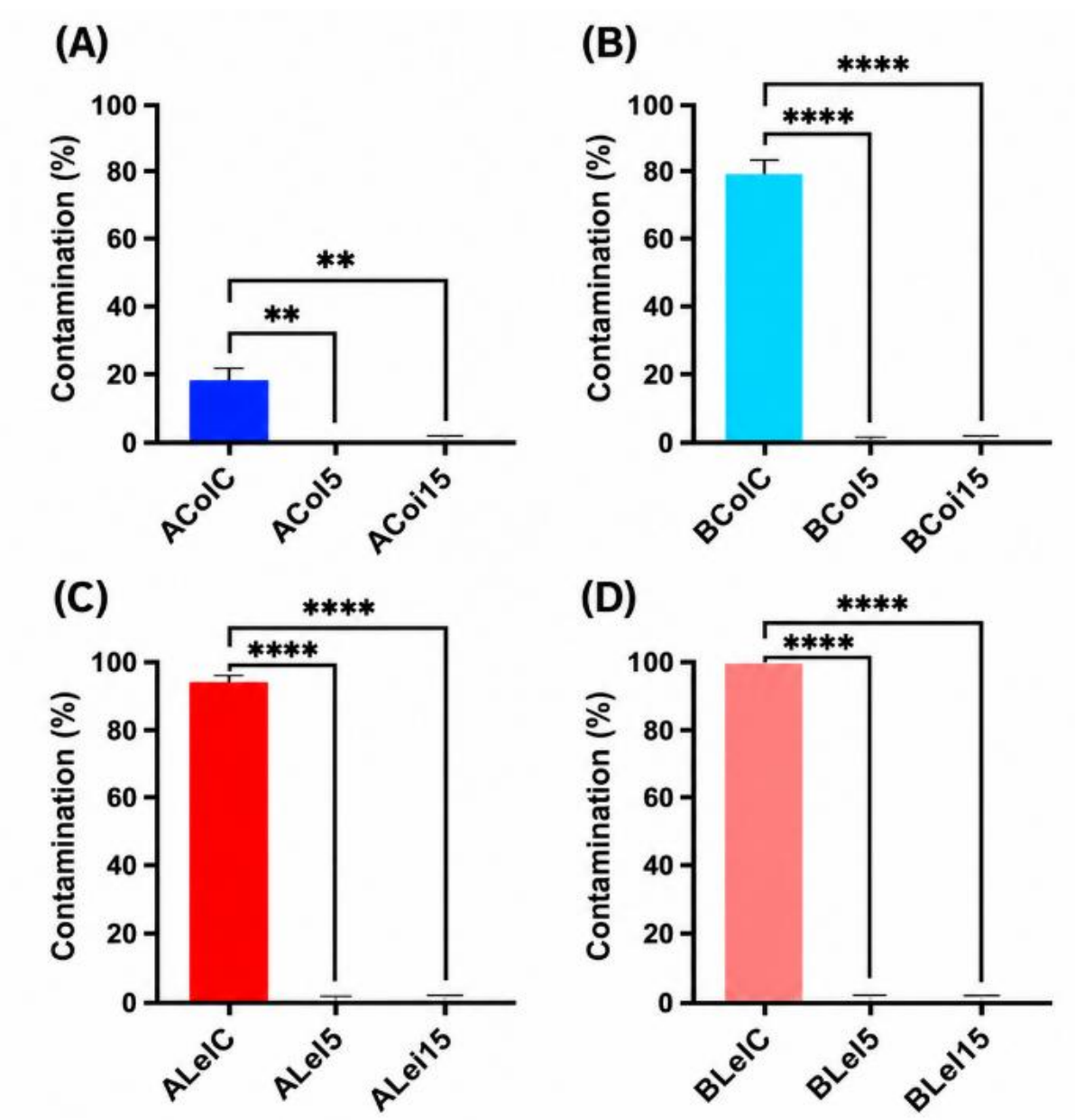

***Figure 1. Effect of CAP treatment on microbial growth from Columbia (Col) and Landsberg (Ler) seeds on malt extract agar (MEA) medium. Seeds were either untreated or plasma-treated for 5 or 15 min. (A) Contamination of Columbia seeds from batch A (ACol*c*, ACol5 and ACol15); (B) contamination of Columbia seeds from batch B (BCol*c*, BCol5 and BCol15); (C) contamination of Landsberg seeds from batch A (ALer*c*, ALer5 and ALer15); (D) contamination of Landsberg seeds from batch B (BLer*c*, BLer5 and BLer15). The percentage of contaminated seeds was determined after 6 days at 21 °C; a seed was considered contaminated when visible microbial growth developed around the seed on the culture medium. Each bar represents the mean of two replicates of 45 seeds ± standard deviation (SD). Statistical analysis was performed using one-way ANOVA followed by Dunnett's multiple comparisons test. Significance levels shown in the figure are: **, p ≤ 0.01; ****, p ≤ 0.0001.***

The morphology of the mycelium after 6 days of growth on MEA medium suggested that the fungus likely belonged to the genus Penicillium (**Supplemental Figure S2**). Macroscopically, Penicillium colonies typically appear as velvety textures with shades of blue-green, often surrounded by a white peripheral margin. In addition, lactophenol blue staining was performed on mycelium and spores directly sampled from the seeds, revealing a characteristic brush-like conidial structure typical of Penicillium species (**Supplemental Figure S2**). A 5 min CAP treatment was sufficient to almost totally suppress seed contamination of all batches (**Figure 1**). After 15 min treatment no contamination at all was detected for any seed batch (**Figure 1**).

#### 2.1.2. CAP treatment inhibits fungal growth assessed by nephelometry

We investigated the effect of CAP treatment on the development of fungal spores in suspension using nephelometry. This experiment was conducted with seeds from Col ecotype from the heavily contaminated batch (batch B, **Figure 1B**). Three hours after inoculation, Nephelometric Turbidity Unit (NTU) values were below 60,000 for the three tested conditions, indicating low initial turbidity in the wells (**Figure 2A**). However, after 48 h, turbidity significantly increased in the untreated condition and in the positive control, indicating fungal growth. In these two conditions, NTU values were 13 and 9 times higher than at the start of the experiment, reaching peak levels of 692,274 and 525,132, respectively. In contrast, in the wells corresponding to 15 min of plasma exposure of seeds, turbidity did not increase between the 3 h and 48 h measurements and remained low (**Figure 2B**). These results, combined with the previous findings (**Figure 1**), indicate that a 15 min CAP treatment effectively prevents the development of fungal spores present on the seeds, thereby reducing the risk of contamination.

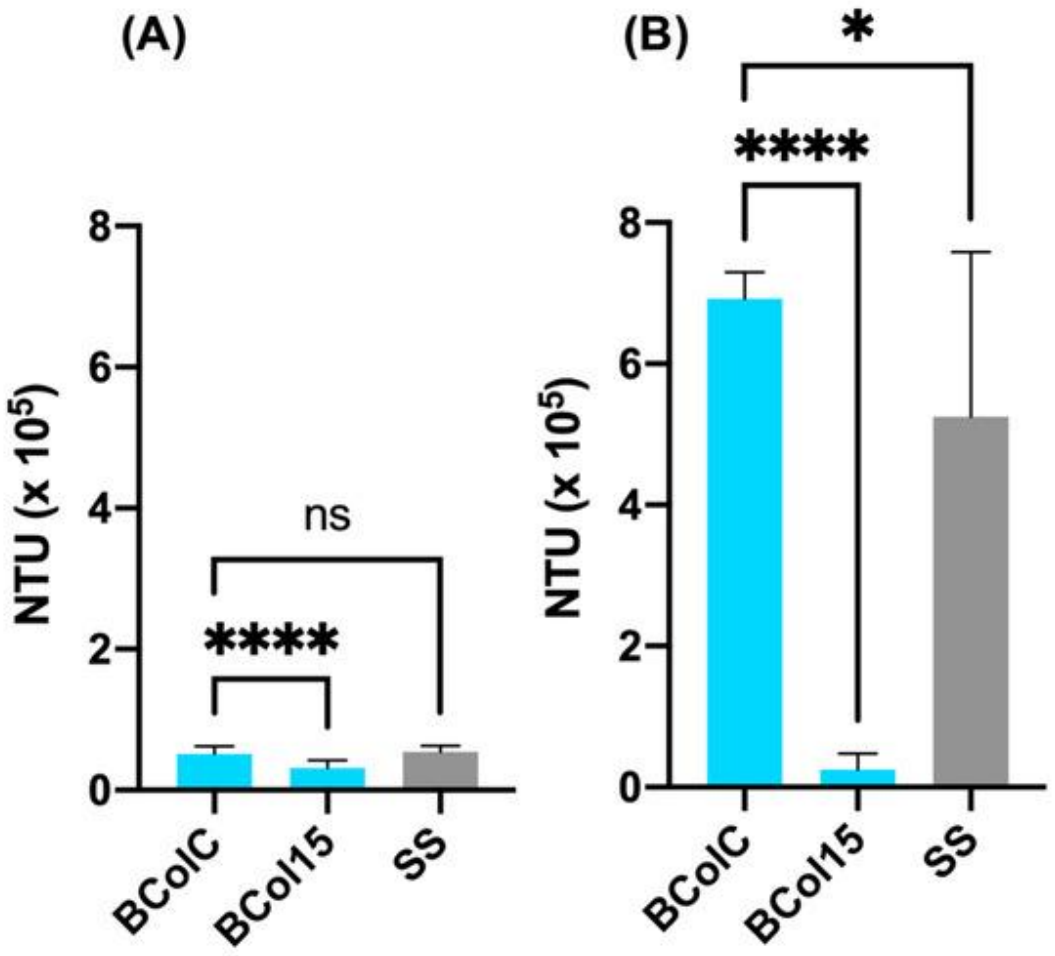

***Figure 2. Fungal growth at 3 h (A) and 48 h (B) after nephelometry measurements. The tested spore suspensions were obtained from untreated (BColC) and plasma-treated (15 min) (BCol15) seeds, and from a spore culture derived from Col seeds (SS, spore suspension), all from the most contaminated batch (batch B). The results correspond to the average NTU (Nephelometric Turbidity Unit) ± standard deviation (SD). Statistical analysis was performed using a one-way ANOVA followed by Dunnett's multiple comparisons test. Significance levels are indicated as follows: ns (p > 0.05), * (p ≤ 0.05), **** (p ≤ 0.0001).***

## 2.2. Deciphering the effects of plasma treatment on seed fungal communities

Amplicon library construction and sequencing were performed on seedlings derived from seeds germinated and grown under sterile conditions. Under these conditions, i.e., on cotton wool moistened with sterile water, no visible bacterial contamination was observed, and only a few cases of visible fungal contamination occurred, exclusively in the control plates, after 8 days. The fungal community associated with seeds contained a small number of dominant taxa common to both ecotypes and both batches, notably Penicillium olsonii, Acremonium sclerotigenum, and Cladosporium cladosporioides (**Figure 3**). Plasma exposure induced significant changes in the composition of fungal communities associated with Col seedlings compared to the untreated condition.

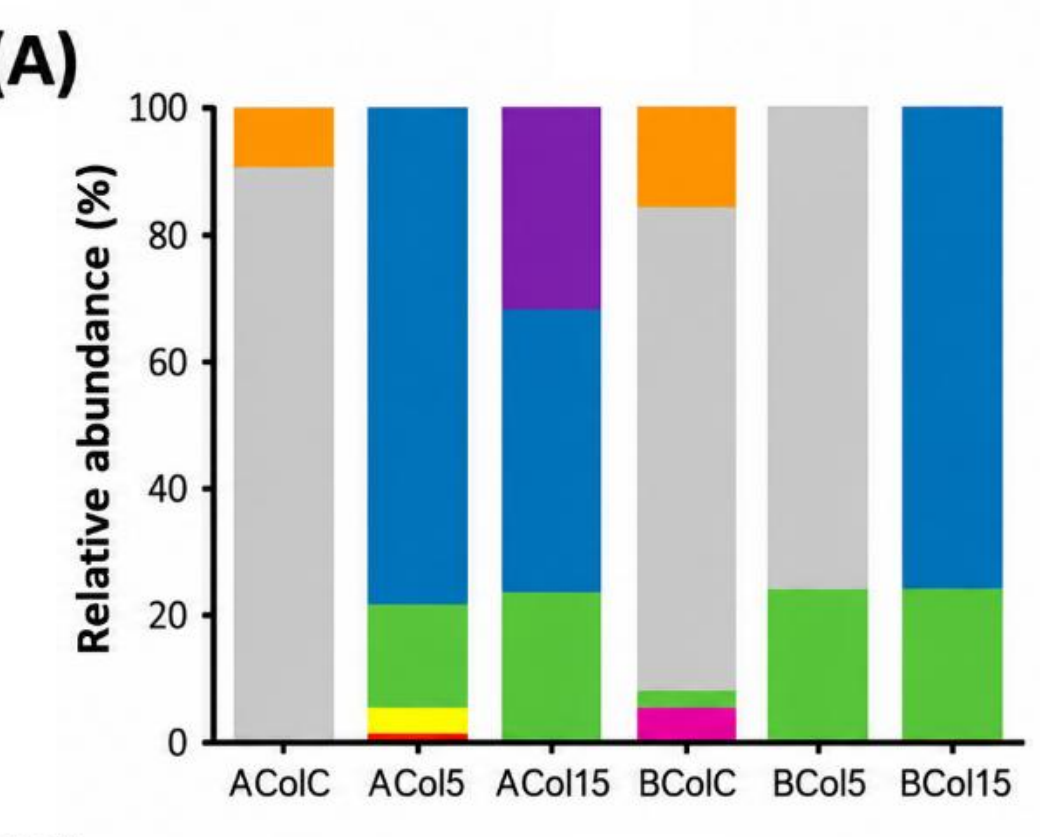

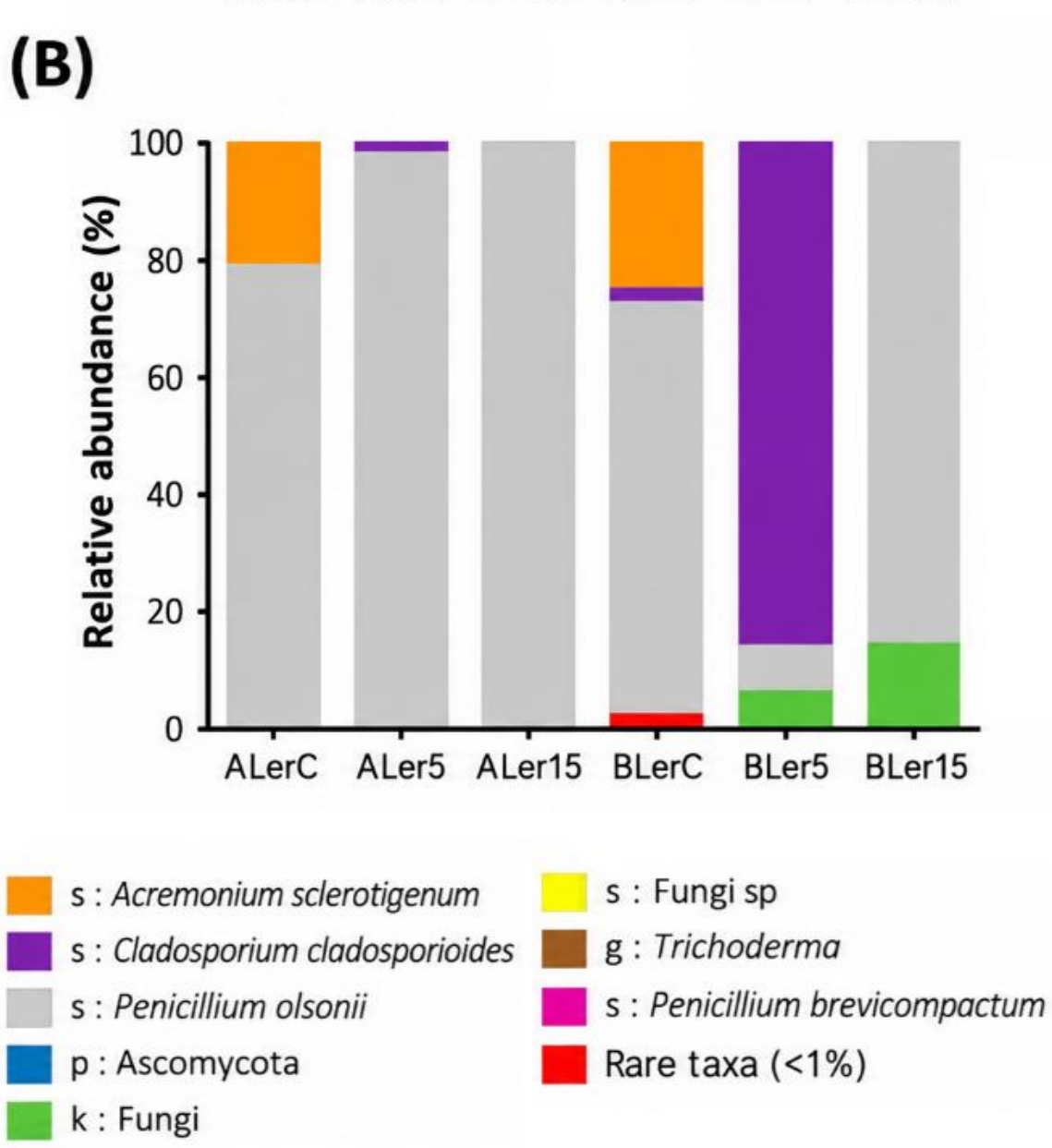

***Figure 3. Taxonomic profiles of fungal communities associated with Arabidopsis seedlings of Columbia (Col) (A) and Landsberg (Ler) (B) ecotypes. Fungal diversity associated with seedlings issued from untreated (C) seeds and seeds treated with plasma for 5- or 15 min from two batches (A & B). Each stacked bar represents the average relative abundance of taxa from 3 samples per condition. Taxa with an abundance greater than 1% are shown in the figure, while the others are grouped under "rare taxa" and are provided as Figure S3.***

The strongest CAP-induced restructuring was observed in Columbia (Col) (**Figure 3A**). In the less contaminated seed batch (batch A), the two dominant taxa of untreated samples, P. olsonii and A. sclerotigenum, became undetectable or represented less than 1% of assigned sequences after CAP treatment. Concomitantly, C. cladosporioides, initially present at less than 1%, represented approximately 30% of sequences after 15 min of treatment. Changes in Col batch B followed the same general pattern but required longer CAP exposure. Penicillium olsonii remained dominant after 5 min of treatment (~75% of sequences), whereas after 15 min its relative abundance decreased below 1%, and approximately 98% of sequences could not be assigned below the phylum level (**Figure 3A**). Thus, although the response differed quantitatively between the two Col batches, prolonged CAP exposure consistently resulted in a strong reduction in the relative representation of the initially dominant fungal taxa. These compositional changes were accompanied by a tendency towards increased alpha diversity in Col samples (**Figure 4A, 4B**). Shannon and Chao1 indices generally increased following CAP exposure, particularly after the strongest restructuring of the dominant community, although differences were not statistically significant. This pattern is consistent with a decrease in the relative dominance of a few abundant taxa, increasing the relative contribution of low-abundance taxa to the community profile, rather than demonstrating an absolute increase in fungal richness.

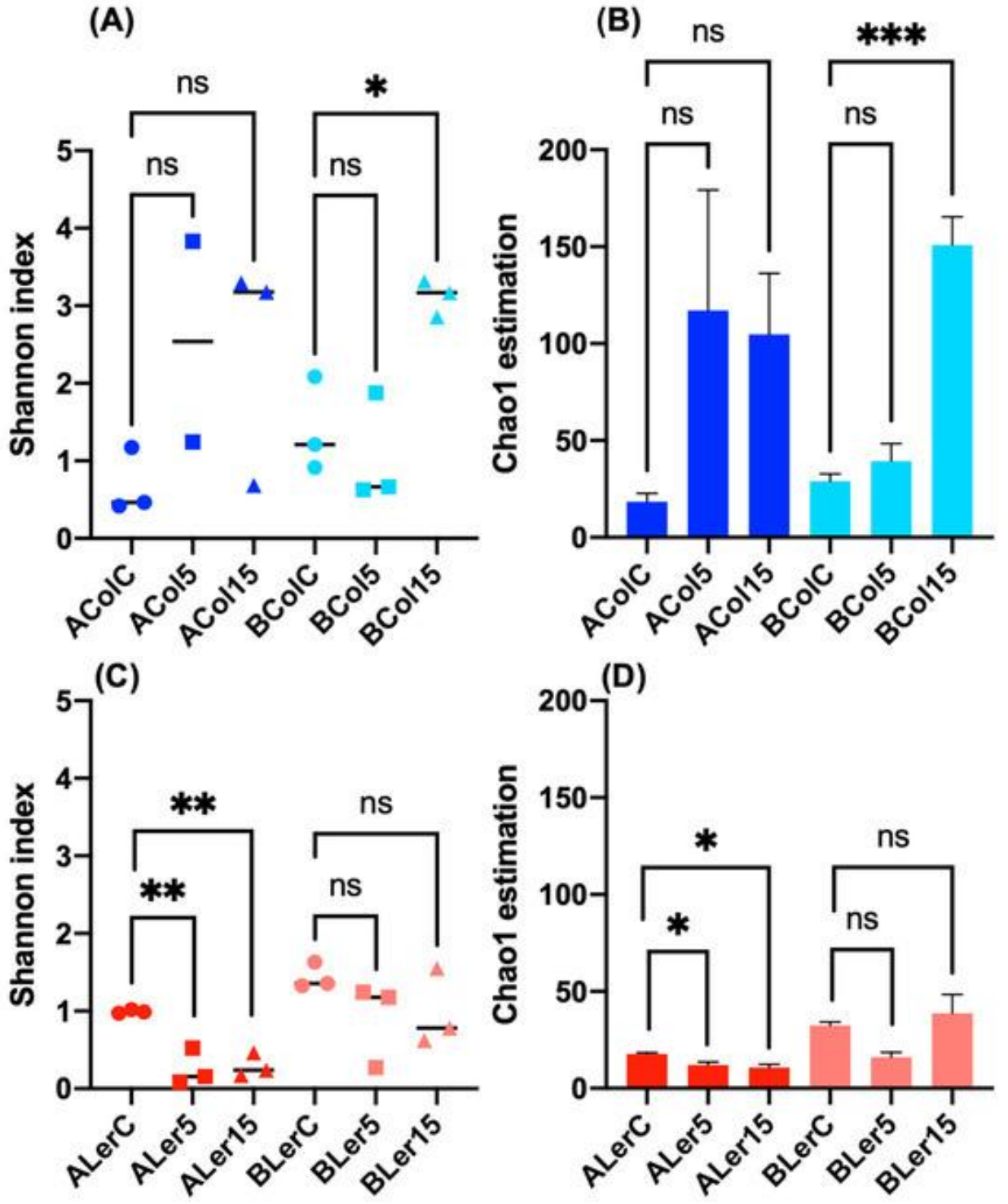

***Figure 4. Alpha diversity metrics of the fungal microbiota (Shannon (A,C) and Chao1 (B,D)) for Columbia (Col) and Landsberg (Ler) seedlings issued from untreated (C) or plasma-treated for 5- or 15 min seeds from batches A and B. Symbols in (A,C) correspond to individual replicates (circles, control; squares, 5 min plasma treatment; triangles, 15 min plasma treatment) and horizontal bars correspond to the means. Statistical analysis was performed using a one-way ANOVA followed by Dunnett's multiple comparisons test. Errors bars represent standard deviation (SD). Significance levels are indicated as follows: ns ($p > 0.05$), * ($p \leq 0.05$), ** ($p \leq 0.01$), *** ($p \leq 0.001$).***

CAP treatment was also associated with changes in the contribution of rare fungal taxa (relative abundance < 1%) in Col seedlings (**Supplemental Figure S3, Supplemental Table S1**). In batch A, rare taxa represented 2.26% of sequences after 5 min of CAP treatment, compared with a very low proportion in untreated samples. In batch B, which initially contained a larger number of rare taxa, their number increased from 8 in untreated samples to 11 following plasma exposure. These changes coincided with a tendency towards higher Shannon and Chao1 indices, although differences were not statistically significant (**Figure 4A, 4B**). Overall, the increased contribution of rare taxa likely reflects the reduced relative dominance of the major fungal taxa following CAP treatment rather than evidence for an absolute increase in fungal diversity.

CAP treatment was associated with a marked increase in sequences that could not be taxonomically resolved beyond the phylum or kingdom level in several samples (A Col5, A Col15, B Col15, **Figure 3A**). Although this increase coincided with the decrease or disappearance of several dominant fungal taxa, the metabarcoding approach does not allow the origin of these poorly assigned sequences to be determined. They were therefore retained as unclassified fungal or Ascomycota sequences without further biological interpretation.

The effect of CAP treatment was also examined by analyzing the total number of reads before rarefaction, which varied across conditions. A significant decrease in the number of reads was observed in treated samples, particularly after 5 min for batch A and after 15 min for batch B (**Supplemental Figure S4**). The observed decrease in the total number of reads in the CAP-treated samples, particularly noticeable after 5 min for batch A and 15 min for batch B, could indicate a reduction in fungal load due to plasma treatment. This reduction in raw read counts may result from the elimination of abundant microorganisms, especially those located on the seed surface and thus directly exposed to plasma. However, because amplicon read numbers can be influenced by DNA extraction, amplification efficiency, library preparation, and sequencing depth, they cannot be used as a direct quantitative measure of fungal load. Consequently, the lower read numbers observed after CAP treatment are reported as an additional sequencing-related observation rather than as evidence of a reduction in fungal biomass.

In contrast to Col, fungal communities associated with Ler seedlings were less consistently affected by CAP treatment (**Figure 3B**). Penicillium olsonii remained highly abundant in most treated samples, particularly in batch A, where it represented more than 97% of sequences following plasma exposure. Consequently, Shannon and Chao1 indices remained low and did not increase following treatment (**Figure 4C, 4D**). Nevertheless, A. sclerotigenum, which represented more than 20% of sequences in untreated Ler samples, decreased to below 1% or became undetectable following CAP treatment in both batches, indicating that individual fungal taxa differed in their response to plasma exposure.

A more pronounced but transient community shift occurred in Ler batch B, where C. cladosporioides increased from 1.3% in untreated samples to 87% after 5 min of CAP treatment, while P. olsonii decreased from approximately 70% to 8% (**Figure 3B**). After 15 min, however, P. olsonii again dominated the community (~85%). Alpha diversity indices did not differ significantly among these conditions (**Figure 4C, 4D**). Rare taxa remained a minor component of the Ler fungal communities across treatments, consistent with the overall low diversity observed in this ecotype (**Supplemental Table S1**). Thus, CAP did not produce a uniform shift in fungal community composition across seed lots.

## 2.3. Fungal beta diversity

Beta diversity was assessed using the Bray–Curtis distance. The Principal Coordinate Analysis (PCoA) revealed a clear separation of fungal communities between the different samples. For the Col ecotype, the first two axes explained 86% of the total variance (PC1 = 72.74%, PC2 = 14.15%) (**Figure 5A**), while for the Ler ecotype, they accounted for 91% (PC1 = 80.7%, PC2 = 10.31%) (**Figure 5B**).

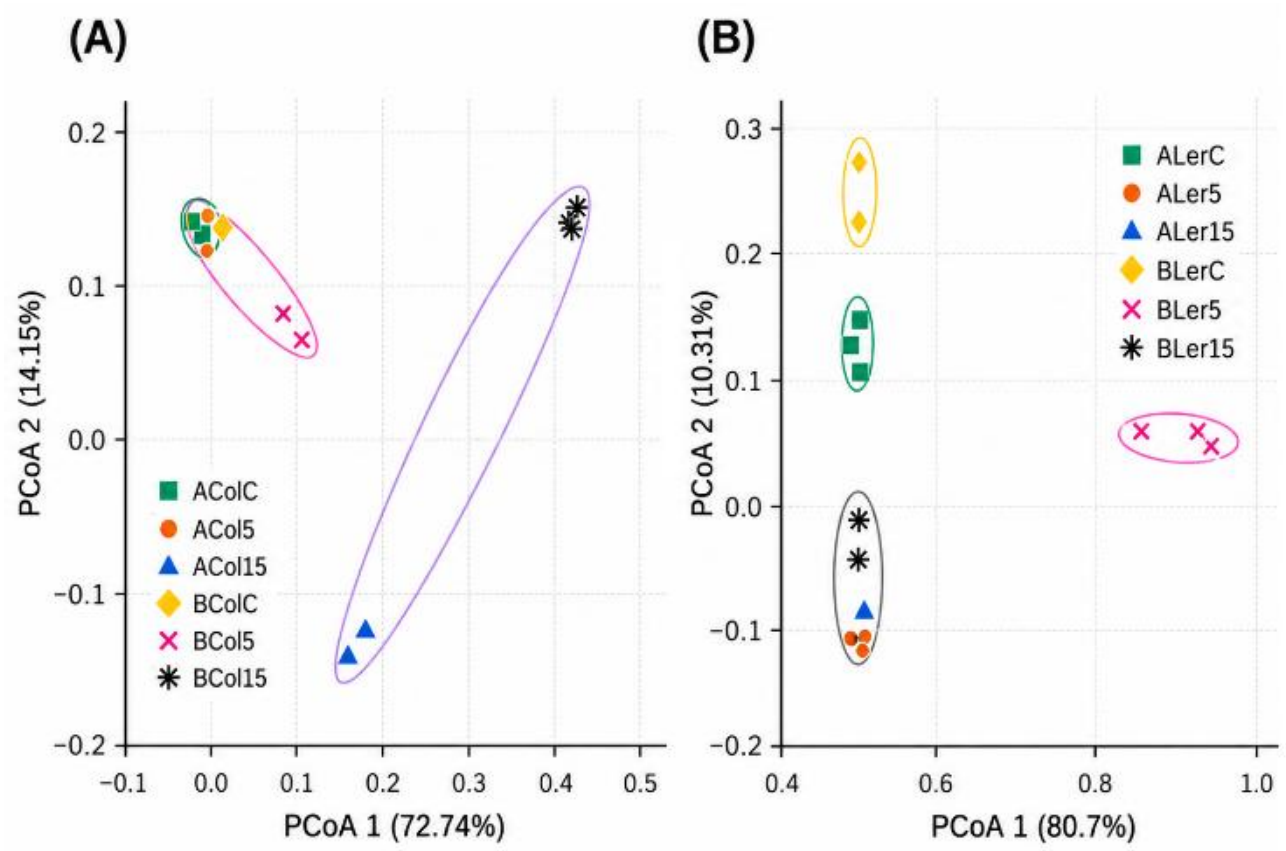

***Figure 5. Principal Coordinate Analysis (PCoA) of fungal communities associated with Columbia (Col) (A) and Landsberg (Ler) (B) seedlings issued from untreated (C) or plasma-treated for 5- or 15 min seeds from batches A and B, based on Bray–Curtis distance matrix analysis of triplicate DNA extractions. Each point represents a biological replicate, and the first two axes illustrate the explained variance in community composition.***

PERMANOVA analysis revealed a significant effect of the group factor, combining seed batch and CAP exposure time, on fungal community composition in both ecotypes ($R^2$ = 77.5% for Col and $R^2$ = 91.5% for Ler; p = 0.001 for both) (**Supplemental Table S2**). In contrast, the batch effect differed between ecotypes: it was not significant in Col ($R^2$ = 4.8%, p = 0.431) but was significant in Ler ($R^2$ = 19.5%, p = 0.004), indicating a greater contribution of seed production conditions to fungal community variation in Ler (**Supplemental Table S2**).

Pairwise PERMANOVA comparisons among individual batch × treatment groups did not remain significant after correction for multiple testing (adjusted p > 0.05), despite high $R^2$ values for some comparisons (**Supplemental Table S3 and S4**). These effect sizes should therefore be interpreted cautiously, as they may partly reflect the limited sample size. When CAP exposure time was considered independently of seed batch, fungal communities differed significantly between treated and untreated seeds in both ecotypes. In Col, significant differences were detected between 5 and 0 min (p = 0.039, $R^2$ = 0.36) and between 15 and 0 min (p = 0.003, $R^2$ = 0.94), with no significant difference between 5 and 15 min (p = 0.052). A similar pattern was observed in Ler, with significant differences between 5 and 0 min (p = 0.048, $R^2$ = 0.33) and between 15 and 0 min (p = 0.012, $R^2$ = 0.70), but not between 5 and 15 min (p = 0.112) (**Supplemental Figure S4B**).

Overall, our findings indicate that CAP treatment was associated with changes in the β-diversity of seed-associated fungal communities, as reflected by Bray–Curtis distances (**Figure 5**). Differences between treated and untreated seeds were detected after 5 min of exposure, while no consistent differences were observed between the 5- and 15 min treatments. In Ler, a significant batch effect also contributed to variation in fungal community composition, indicating that seed production conditions influenced β-diversity patterns. Although some pairwise comparisons yielded high $R^2$ values, they did not remain statistically significant after correction for multiple testing and therefore cannot be considered as evidence of confirmed differences between groups. These results should consequently be interpreted with caution.

# 3. Discussion

In this study, we demonstrate that short cold atmospheric plasma (CAP) treatments efficiently decontaminate Arabidopsis thaliana seeds and significantly influence the structure of seedling-associated fungal communities, without altering germinative properties. The data obtained from assessing seed contamination levels, either on MAE medium (**Figure 1**) or by nephelometry (**Figure 2**), demonstrate that 15 min of CAP treatment reduced contamination to undetectable levels in seeds of both ecotypes, regardless of their initial contamination level, including the most heavily contaminated batches. The effects of plasma treatments on seed contamination are well known and have been evidenced in a wide range of species for fungi and bacteria [21], [22]. Although not fully understood, it has been proposed that this effect might mainly result from reactive oxygen species (ROS) and reactive nitrogen species (RNS) produced by the plasma, which can in turn lead to various oxidative damage towards microorganisms [22].

Surprisingly, the Arabidopsis seed microbiota has been poorly studied until now, the studies by Barret et al. [23] and Johnston-Monje et al. [24] being among the few investigating fungal and bacterial communities in seeds of sf-2 and Col ecotypes, respectively. In their study, Johnston-Monje et al. [24] showed that the fungal microbiome of Col seeds was dominated by the phylum Ascomycota and bacteria were almost all proteobacteria. The effects of cold plasma on seed-associated fungal communities have also been investigated in buckwheat, where plasma treatment altered fungal community structure and different fungal taxa displayed contrasting sensitivities to treatment [20]. Here, as stated before, we investigated the microbiota of 7 d old seedlings issued from seeds that were or were not treated with plasma exposure. As shown by Barret et al. [23], the process of emergence significantly impacts the structure of microbiota, leading to a decrease in both bacterial and fungal diversity; thus, we only provide a proxy of microbiota in seeds. Here Penicillium olsonni and Acremonium sclortigenum were the most represented fungi in all seedlings, independently of the ecotype and of the production (**Figure 3**).

The effect of plasma on fungal communities greatly depended on the ecotype and on the seed batch, as first shown by the number of reads in each sample, which decreased with plasma treatment in Col seedlings whereas it remained unaffected in Ler seedlings. The contrasting responses of Col and Ler therefore indicate that the effect of CAP on seed-associated fungal communities is context-dependent. The greater restructuring observed in Col than in Ler may reflect differences associated with host genotype, but also differences in the initial composition and abundance of the seed-associated microbiota. The batch-dependent responses within each ecotype further support an important contribution of the initial microbial community. The present experimental design does not allow host-genotype effects to be separated from initial microbiota effects, but it demonstrates that CAP efficacy against individual fungal taxa depends on the biological context of the seed lot. The observed decrease in the total number of reads in Col plasma-treated samples, which was particularly noticeable after 5 min for batch A and 15 min for batch B, could indicate a reduction in fungal load due to plasma treatment. This reduction in raw read counts may result from the elimination of abundant microorganisms, especially those located on the seed surface and thus directly exposed to plasma. This suggests that the treatment may have led to a decrease in overall fungal abundance in the Col treated samples, but not in the Ler treated ones.

In all samples, plasma exposure resulted in the disappearance of Acremonium sclerotigenum DNA, while Penicillium olsonii DNA remained dominant in Ler seedlings, regardless of the treatment duration (**Figure 3B**). However, since no fungal growth was detected on Col and Ler seeds exposed to plasma (**Figure 1**), we hypothesize that the detection of Penicillium olsonii DNA in seedlings derived from plasma-treated seeds reflects the presence of DNA from non-viable fungi. In some CAP-treated samples (A Col5, A Col15, B Col15, and B Ler15; **Figure 3**), a significant proportion of sequences could only be assigned to Ascomycota or Fungi. One possible explanation is that plasma-induced damage to fungal material, including DNA damage, may have reduced the recovery of intact and reliably classifiable fungal amplicons. However, this interpretation remains hypothetical, since the metabarcoding approach used here cannot distinguish DNA degradation from other factors affecting taxonomic assignment, including sequence quality, amplification biases, or limitations of the reference database. Therefore, the increase in poorly assigned sequences should not be considered direct evidence of plasma-induced fungal DNA degradation.

In certain cases, plasma treatment was associated with the development of novel fungal communities, mostly rare species, in Col seedlings (**Figure 3A**), which was reflected by an increase in Shannon and Chao1 indexes (**Figure 4A, 4B**). We propose that the disappearance of the dominant and viable Penicillium olsonii allowed the colonization of seedling tissues by other fungi, notably endophytes present in low abundance, once Penicillium olsonii lost its competitive advantage. This was not the case for Ler seedlings, where diversity indices remained low before and after plasma treatment, with very few abundant or rare taxa. Only the relative abundance of the genus Trichoderma, found exclusively on Ler seedlings and initially below 1% in samples from batches A and B, was found at higher relative abundance in the samples from batch B issued from seeds exposed for 5 min. It is worth noting that several endophytic fungi belonging to this genus are being studied for their potential to stimulate plant growth due to their ability to solubilize phosphorus present in the soil **[25]**, as well as to enhance plant stress resistance, notably drought tolerance **[26]**. Finally, our statistical analysis identified exposure time as the primary driver of fungal community variation in both ecotypes. Interestingly, our findings indicate that fungal species exhibited different susceptibility, resilience and colonization potential thresholds to treatments. Upon treatment, Cladosporium cladosporioides appeared more opportunistic and resistant, whereas Acremonium sclerotigenum systematically declined in relative abundance in exposed samples, even after 5 min of treatment.

Beyond its decontamination efficiency, CAP treatment may have important ecological implications because it modifies the pool of microorganisms available to colonize developing seedlings. In the present study, the elimination of dominant seed-borne fungal taxa, particularly Penicillium, was associated with shifts in the fungal communities detected after germination, suggesting that plasma treatment can alter early microbial assembly processes. Such changes may result from the release of ecological niches previously occupied by sensitive microorganisms, thereby enabling the proliferation of taxa that persist after treatment or originate from internal seed tissues. From an agricultural perspective, these findings highlight that CAP should not be viewed solely as a tool for pathogen control but also as a factor capable of influencing plant-associated microbiota. Because seed-associated microorganisms can contribute to plant growth, stress tolerance and disease suppression, understanding how plasma treatment affects microbial community assembly will be important for optimizing treatment conditions. Future work should therefore aim to identify treatment parameters that effectively reduce undesirable microorganisms while preserving or promoting beneficial members of the seed microbiome.

This study confirms the effectiveness of a short cold atmospheric plasma (CAP) treatment for fungal decontamination of Arabidopsis thaliana seeds. CAP strongly reduced the abundance of seed-associated fungi and markedly altered the composition of fungal communities detected in developing seedlings. In particular, the treatment eliminated dominant seed-borne fungal taxa, including Penicillium, and was associated with shifts in the relative abundance of other fungal groups during seedling establishment. Our results further demonstrate that the impact of CAP treatment on seedling-associated fungal communities is influenced by both seed lot and plant genotype, highlighting the complexity of plant–microbe interactions following microbiome perturbation. Together, these findings show that CAP is not only an efficient seed decontamination technology but also a factor capable of modifying the fungal microbiota associated with emerging seedlings. Future studies should investigate the mechanisms underlying microbial persistence or elimination following plasma exposure and assess the long-term consequences of these microbiome changes for plant development and health. These analyses should also be extended to agriculturally important crop species and larger, naturally contaminated seed lots to determine whether the microbiota changes observed here are maintained under more representative production conditions. Such validation, together with the development or evaluation of plasma systems suitable for high-throughput seed treatment, will be important for assessing the potential transfer of CAP technology from laboratory experiments to agricultural applications.

# 4. Materials and methods

## 4.1. Plant material and germination assays

This study was performed with two ecotypes of Arabidopsis thaliana: Columbia-0 (Col) and Landsberg-0 (Ler). Seeds used in this study were produced in-house under controlled growth conditions. Seeds were stratified for 4 d at 4 °C before being sown in soil and placed in growth chambers under non-sterile conditions with a temperature of 20–22 °C of and a photoperiod of 16 h per day with the light intensity of 150 µmol $m^2$ $s^1$. Seeds were collected from mature siliques after 2 weeks drying at room temperature and subsequently stored at room temperature. Two batches of seeds (A and B) from two independent production cycles were used for each ecotype. Seeds were germinated at 15 °C on a layer of cotton wool covered by a filter paper sheet soaked with water for 10 d at 15 °C in darkness.

A seed was considered germinated when the radicle had protruded through the testa. The results presented correspond to the mean of the germination percentages obtained for 3 replicates of 50 seeds.

## 4.2. Plasma treatment

Dry seeds were treated with cold atmospheric plasma (CAP) using a dielectric barrier discharge (DBD) system [27], [28]. Air plasma was generated in a DBD device consisting of two flat electrodes, each measuring 30 × 40 $cm^2$, one connected to a high voltage source and the other grounded. The electrodes were separated by a 2 mm thick insulating dielectric material to prevent any transition from streamer to arc, ensuring uniform plasma distribution within the reactor where the seeds were exposed. The background gas used was ambient air, with a relative humidity of approximately 40%. The high-voltage electrode was powered by a 9 kV supply at a frequency of 140 Hz (sine wave), provided by a high-voltage generator comprising a function generator (ELC, Annecy, France, GF467AF) and a power amplifier (CC5500, Crest Audio, Meridian, MS, USA). The two durations of treatment tested here, 5 min and 15 min, were selected based on previous studies [28] and preliminary experiments indicating that they represent distinct levels of plasma exposure while maintaining seed viability. The inter-electrode gap was 1 mm.

## 4.3. Assessment of seed contamination on malt extract agar medium

For each replicate, between 40 and 50 non-disinfected seeds for each replicate were placed under sterile conditions onto malt extract agar (MEA) medium (30 g malt extract and 10 g agar per liter) for each untreated and plasma-treated condition (5 and 15 min of plasma treatment) and each ecotype (Col and Ler) in 2 replicates (each consisting of at least 40 seeds). Photos were taken daily to monitor contamination levels, and the percentage of contaminated seeds was calculated after 6 days at 21 °C with a 16 h light/8 h dark photoperiod. A seed was considered contaminated when visible microbial growth developed around the seed on the culture medium during incubation. One-way Anova followed by Dunnett's multiple comparisons tests were performed using Prism 9 to compare results across conditions and GraphPad Prism 9 was used to generate figures.

## 4.4. Nephelometry

Nephelometry was performed on Columbia seeds from the highly contaminated batch B. Eighty mg of Col seeds from each treatment condition (untreated control, 5 min and 15 min CAP treatment) were processed separately. Seeds from each condition were stirred for 1 h in sterile liquid malt medium (30 g malt extract L−1) to recover epiphytic microorganisms. The incubation mixture from each sample was then filtered to retain the seeds and collect the medium containing the detached fungal spores. A volume of 200 µL per well were pipetted under sterile conditions into a 96-well plate. On the plate, 12 wells were filled per condition, changing rows every 4 wells to prevent cross-contamination between conditions. A negative control (sterile liquid malt medium) and a positive control, consisting of a spore suspension prepared in sterile liquid malt medium from a Penicillium culture isolated from control seeds, were included. To prepare the spore suspension, the spores were scraped off the mycelium grown on solid malt agar mixed with 100 µL of liquid malt medium and filtered using filter cloth (10 µm pores) and diluted in sterile liquid malt medium. The concentration of the 3 control suspensions was measured with a Malassez counting chamber (Jeulin, Evreux, France) and adjusted by adding liquid malt medium to reach final concentrations of 103, 104, and 105 spores/mL. Nephelometric assays were conducted using a NEPHELOstar® Plus microplate reader (BMG LABTECH GmbH, Ortenberg, Germany), equipped with a 635 nm laser diode capable of detecting scattered light up to an 80° angle. Growth was automatically recorded for 72 h at 25 °C and measurements were done every 20 min with a laser intensity of 40%. The data are given in nephelometric turbidity units (NTU) and were analyzed using MARS software (version 3.31).

## 4.5. Metabarcoding analysis

Seeds were sown under sterile conditions on moistened cotton pads with sterile water, in two replicates. Both water and cotton were sterilized. After 8 d of incubation at 21 °C with a 16 h light/8 h dark photoperiod, 0.1 g of seedlings (i.e., approx. 35 seedlings) were collected per condition (control, 5 and 15 min) and per ecotype (Col and Ler) in triplicates. Whole Arabidopsis seedlings were harvested, immediately frozen and stored at −80 °C until analysis. Since DNA was extracted from whole 8-day-old seedlings, the detected microbial DNA may originate both from microorganisms actively associated with seedling tissues and from residual DNA derived from seed-associated microorganisms that persisted after germination. Prior to DNA extraction, approximately 100 mg of seedlings was ground to a fine powder in a pre-chilled mortar using liquid nitrogen. The resulting powder was then processed for genomic DNA extraction using the DNeasy Plant Pro Kit (QIAGEN, Hilden, Germany) according to the manufacturer's instructions. The extracted DNA samples were sequenced by Novogene Europe (Cambridge, UK) using paired-end Illumina (San Diego, CA, USA) amplicon sequencing, targeting the ITS1 region of the internal transcribed spacer for fungi [29], [30]. Primer sequences are given in **Supplemental Table S5**. Sequencing depths of 50k reads for the ITS1 region were applied.

## 4.6. Bioinformatic analyses

Raw paired-end sequencing reads were first demultiplexed according to their unique barcodes, and barcode and primer sequences were removed. Paired-end reads were then merged using FLASH (v 1.2.11) to generate contiguous sequences, referred to as raw tags. Quality filtering was subsequently performed following the QIIME pipeline to remove low-quality sequences and obtain high-quality clean tags. Chimeric sequences were identified by comparison with the Gold reference database using the UCHIME algorithm and removed, resulting in the final set of effective tags. All downstream analyses were conducted using the clean data. The amplicon was sequenced on Illumina paired-end platform to generate 250 bp paired-end raw reads (Raw PE) and then merged and pre-treated to obtain Clean Tags. The chimeric sequences in Clean Tags were detected and removed to obtain the Effective Tags which can be used for subsequent analysis. The main characteristics of sequencing obtained in each step of data processing is shown **Supplemental Table S6**.

The clean sequencing data for fungal ITS regions was analyzed using the QIIME2 (version 2024.5) bioinformatics platform, leveraging GitHub resources from Simonin et al. (2022) [29]. Taxonomic classification of each amplicon sequence variant (ASV) was performed using the naïve Bayes classifier implemented via the scikit-learn package (version 1.9.0). in QIIME2, with the UNITE release 10 database used as reference for ITS sequences. Sequence data analyses were primarily conducted using QIIME2 and R packages (R version 4.5.0 and RStudio version 2024.12). The total number of ASVs obtained after the full processing pipeline is given as **Supplemental Data S1**. Alpha diversity indices at the ASV level, including the Chao1 richness estimator, Shannon index, and Pielou's evenness, were calculated in QIIME2. Principal coordinate analysis (PCoA) was performed in R to estimate and visualize beta diversity based on the Bray–Curtis distance matrix generated in QIIME2. PERMANOVAs and pairwise PERMANOVAs were conducted in RStudio to assess the statistical significance ($\alpha = 0.05$; 999 permutations used for each test) of differences between groups, with Bonferroni correction applied for multiple comparisons. Differences in the number of contaminated seeds, diversity indices, and read counts prior to rarefaction were evaluated in Prism 9 using unpaired t-tests. Rarefaction thresholds were determined based on rarefaction curves to maximize diversity while retaining the highest number of samples in the analysis. A threshold of 51,364 reads for fungal communities was applied. One replicate from the Columbia batch A exposed to CAP for 5 min was excluded because it failed the predefined rarefaction threshold and therefore could not be retained for downstream diversity analyses. This exclusion did not alter the overall conclusions of the study.

# 5. Supplementary materials

The following supporting information can be downloaded at: https://www.mdpi.com/article/10.3390/plants15172719/s1, Table S1: Alpha diversity indexes for Columbia and Landsberg based on ITS region; Table S2: PERMANOVA results for group and batch factors in Columbia and Landsberg; Table S3: Pairwise PERMANOVA results for group (a) and time (b) factors in Columbia; Table S4: Pairwise PERMANOVA results for group (a) and time (b) factors in Landsberg; Table S5: IST1-1F primer used in this study; Table S6: Summarizations obtained in each step of processing for sequencing data of fungal ITS; Data S1: Number of ASVs after sequencing for fungal ITS region; Figure S1: Germination of CAP treated seeds; Figure S2: Microscopic observations of Penicillium structures; Figure S3: Taxonomic profiles of rare taxa of fungal communities; Figure S4: ITS1 region read counts for Columbia (a) and Landsberg (b) in seedlings.

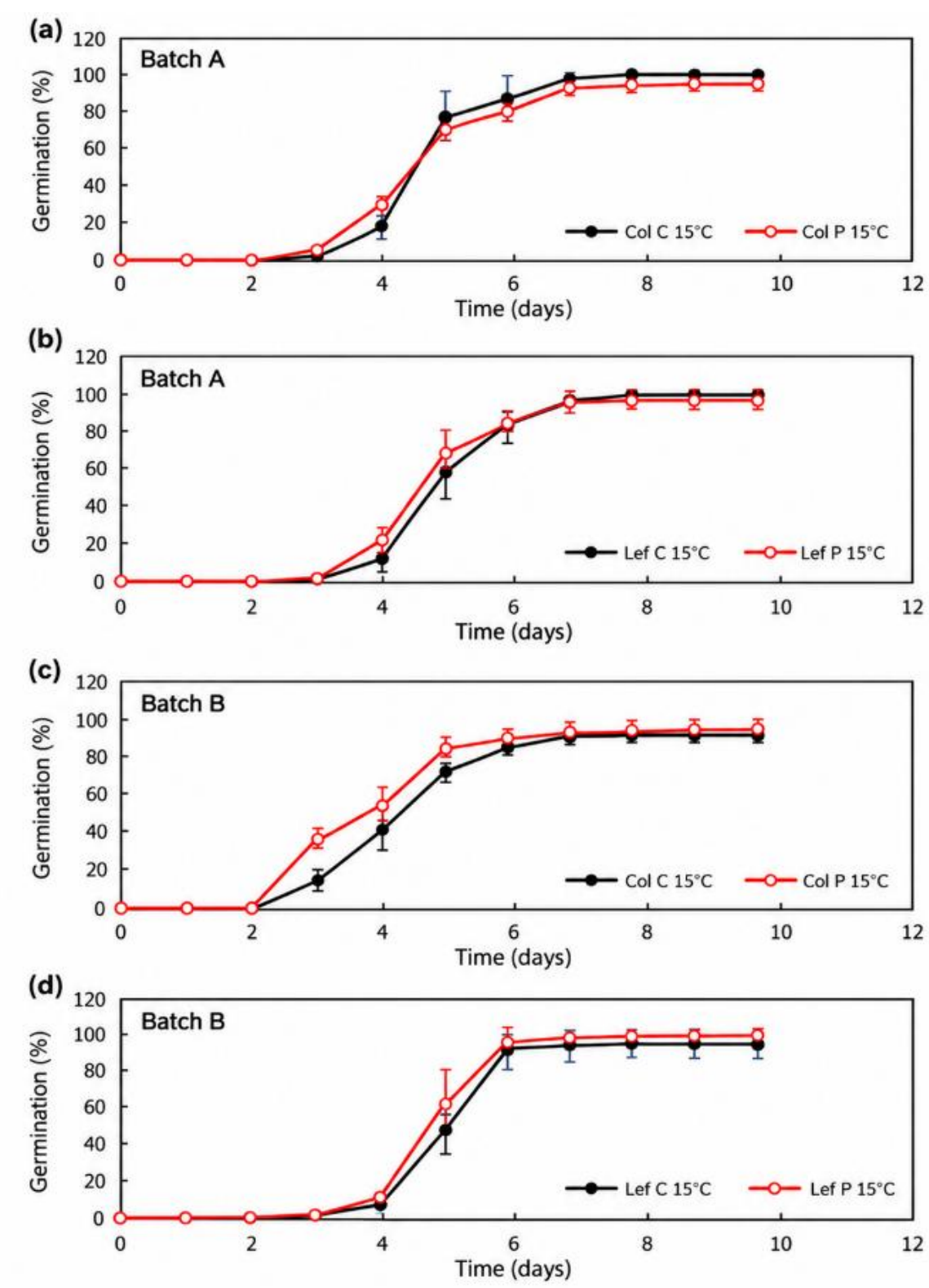

***Figure S1. Germination at 15°C of Arabidopsis seeds treated (P) or not (C) by 15 min Cold Atmospheric Plasma. Ecotypes Col (a,c) and Ler (b,c) from 2 batches (A,B) indicated in the figures. Means +/- SD of 3 replicates of 50 seeds.***

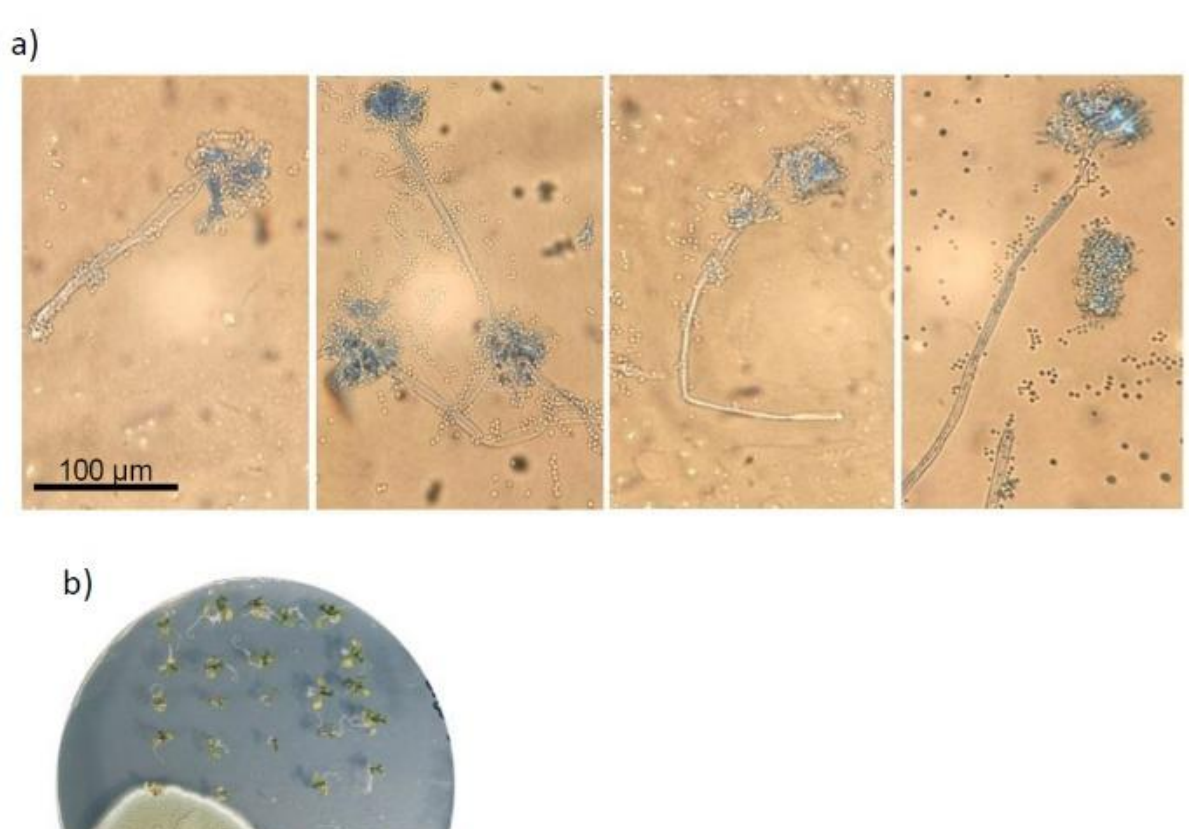

***Figure S2: (a) Microscopic observations (×64) of Penicillium structures isolated from Arabidopsis seeds after lactophenol blue staining, showing characteristic brush-like spore arrangements. b: Colony morphology of Penicillium on MS/2 Medium.***

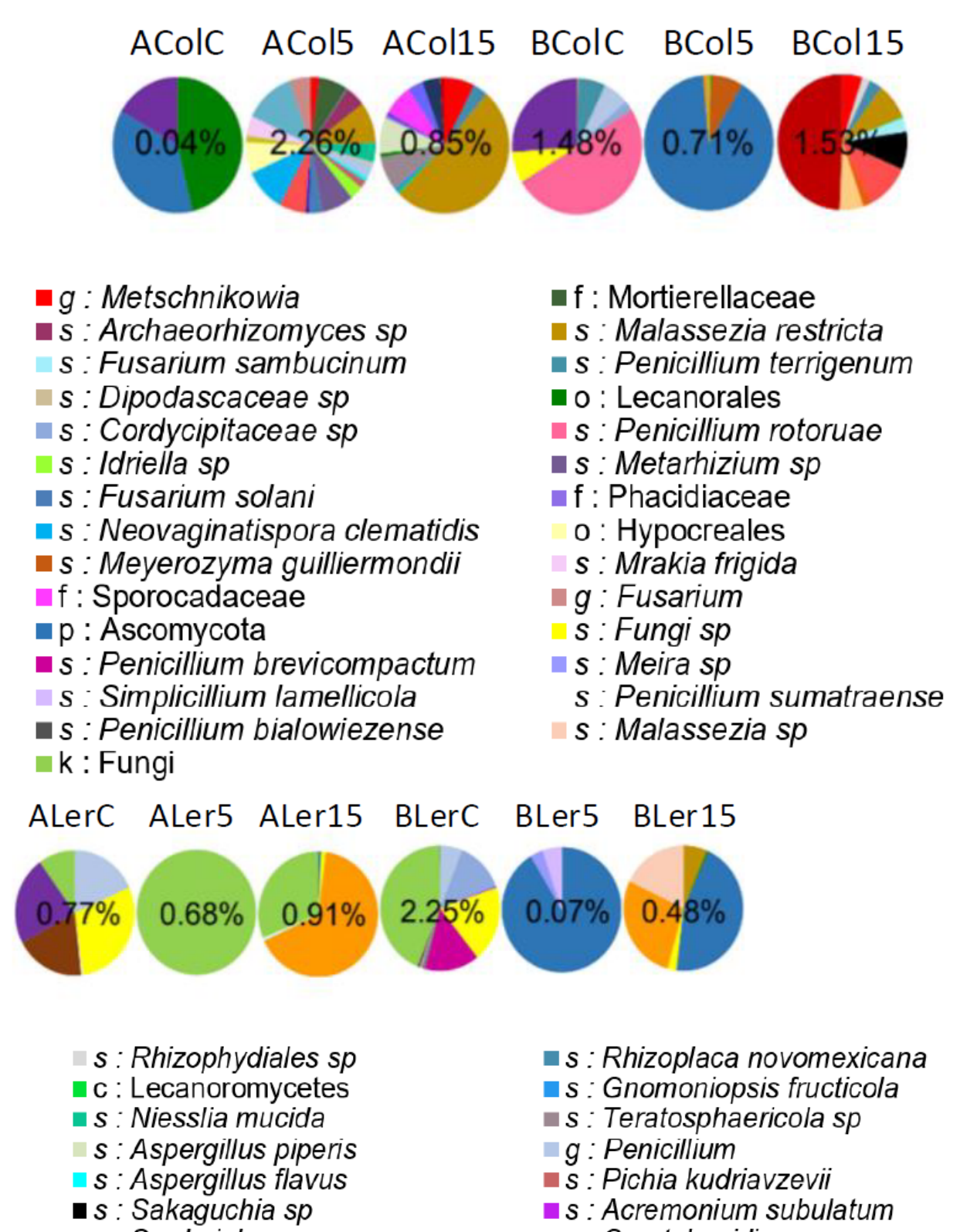

***Figure S3: Taxonomic profiles of rare taxa of fungal communities associated with Arabidopsis seedlings of Columbia (Col) and Landsberg (Ler) ecotypes. Fungal diversity associated with seedlings issued from untreated (C) and plasma-treated for 5- or 15- min seeds from two batches (A & B). Each pie chart represents the average relative abundance of rare taxa from 3 samples per condition.***

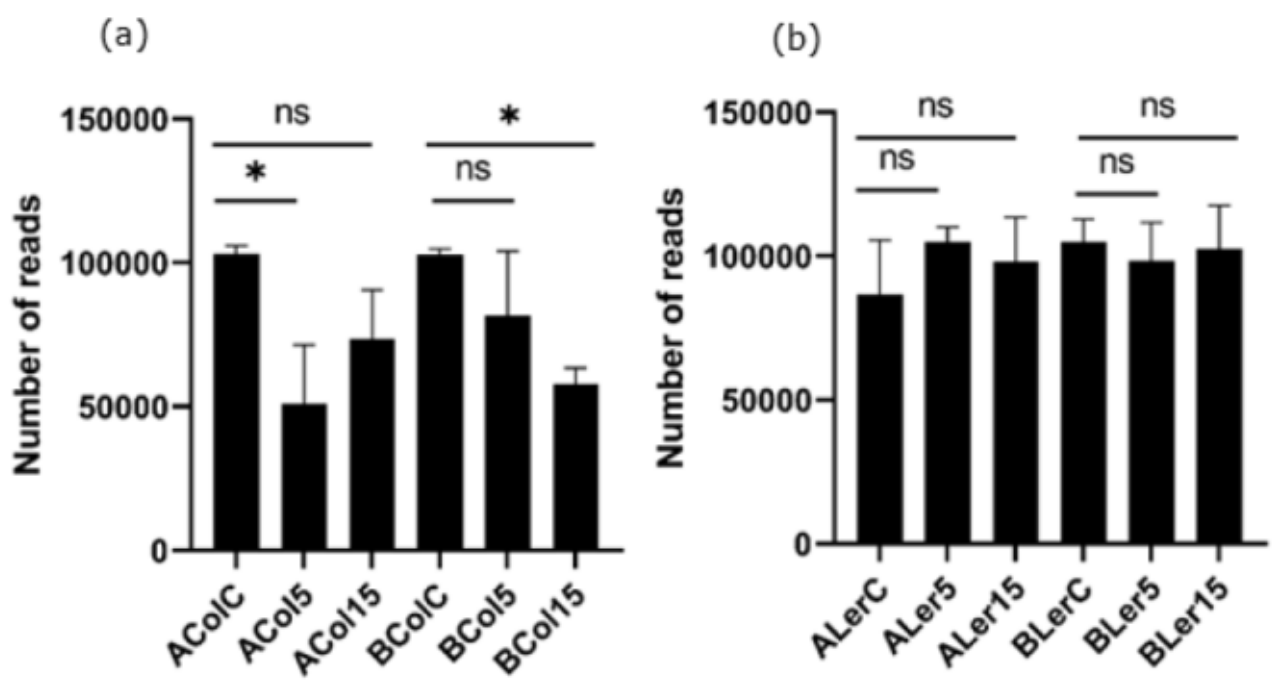

***Figure S4: ITS1 region read counts for Columbia (a) and Landsberg (b) in seedlings issued from untreated (C) or plasma-treated for 5- or 15- min seeds from batches A and B. Statistical significance was assessed using an unpaired t-test. Significance levels are indicated as follows: ns ($p$ > 0.05), * ($p \leq 0.05$), ** ($p \leq 0.01$), *** ($p \leq 0.001$), **** ($p \leq 0.0001$).***

| Sample | Shannon | Pielou | Chao1 | Observed features | Goods coverage |
|---|---|---|---|---|---|
| AColC | 0.68521836 | 0.16577108 | 18.3809524 | 18.333 | 0.999 |
| ACol5 | 2.53814939 | 0.36365919 | 117.166667 | 117 | 0.999 |
| ACol15 | 2.38571795 | 0.34703811 | 104.733333 | 104.333 | 0.999 |
| BColC | 1.40402032 | 0.28718427 | 29 | 29 | 1 |
| BCol5 | 1.05548656 | 0.19452278 | 39.4166667 | 39.333 | 0.999 |
| BCol15 | 3.11323685 | 0.43121275 | 150.833333 | 150.333 | 0.999 |
| ALerC | 0.99296607 | 0.23988716 | 17.6666667 | 17.666 | 0.999 |
| ALer5 | 0.25434314 | 0.06963088 | 12 | 12 | 0.999 |
| ALer15 | 0.29343045 | 0.09114556 | 10.6666667 | 10.666 | 0.999 |
| BLerC | 1.43811234 | 0.28656182 | 32.3333333 | 32.333 | 0.999 |
| BLer5 | 0.89870775 | 0.23608469 | 16 | 16 | 0.992 |
| BLer15 | 0.98531905 | 0.18454257 | 38.6666667 | 38.666 | 0.999 |

***Table S1 : Alpha diversity indexes for Columbia and Landsberg based on ITS region***

| Ecotype | Factor | Df | Sum of Squares | $R^2$ (%) | F-value | p-value | Significativity |
|---|---|---|---|---|---|---|---|
| Columbia | Group | 5 | 3.7942 | 77.5 | 7.5741 | 0.001 | *** |
| Columbia | Batch | 1 | 0.2331 | 4.8 | 0.7497 | 0.431 | ns |
| Landsberg | Group | 5 | 2.20493 | 91.5 | 25.706 | 0.001 | *** |
| Landsberg | Batch | 1 | 0.47041 | 19.5 | 3.879 | 0.004 | ** |

***Table S2 : PERMANOVA results for group and batch factors in Columbia and Landsberg.***

(A)

| Group | Pairs | | Df | SumsOfSqs | F.Model | R2 | p.value | p.adjusted |
|---|---|---|---|---|---|---|---|---|
| 1 | ACol-treated15 vs | ACol-treated5 | 1 | 0.006 | 0.512 | 0.14 | 1.0 | 1.00 |
| 2 | ACol-treated15 vs | ACol-untreated | 1 | 0.370 | 39.505 | 0.90 | 0.1 | 0.13 |
| 3 | ACol-treated15 vs | BCol-treated15 | 1 | 0.013 | 1.566 | 0.28 | 0.1 | 0.13 |
| 4 | ACol-treated15 vs | BCol-treated5 | 1 | 0.272 | 16.919 | 0.80 | 0.1 | 0.13 |
| 5 | ACol-treated15 vs | BCol-untreated | 1 | 0.337 | 34.886 | 0.89 | 0.1 | 0.13 |
| 6 | ACol-treated5 vs | ACol-untreated | 1 | 0.387 | 328.121 | 0.99 | 0.1 | 0.13 |
| 7 | ACol-treated5 vs | BCol-treated15 | 1 | 0.002 | 4.468 | 0.59 | 0.1 | 0.13 |
| 8 | ACol-treated5 vs | BCol-treated5 | 1 | 0.2943 | 28.911 | 0.90 | 0.1 | 0.13 |
| 9 | ACol-treated5 vs | BCol-untreated | 1 | 0.356 | 223.219 | 0.98 | 0.1 | 0.13 |
| 10 | ACol-untreated vs | BCol-treated15 | 1 | 0.551 | 750.626 | 0.99 | 0.1 | 0.13 |
| 11 | ACol-untreated vs | BCol-treated5 | 1 | 0.009 | 1.145 | 0.22 | 0.3 | 0.34 |
| 12 | ACol-untreated vs | BCol-untreated | 1 | 0.002 | 1.384 | 0.25 | 0.2 | 0.25 |
| 13 | BCol-treated15 vs | BCol-treated5 | 1 | 0.422 | 56.385 | 0.93 | 0.1 | 0.13 |
| 14 | BCol-treated15 vs | BCol-untreated | 1 | 0.5068 | 485.464 | 0.99 | 0.1 | 0.13 |
| 15 | BCol-treated5 vs | BCol-untreated | 1 | 0.008 | 1.005 | 0.20 | 0.5 | 0.53 |

(B)

| Time | Pairs | Df | SumsOfSqs | F.Model | R2 | p.value | p.adjusted | sig |
|---|---|---|---|---|---|---|---|---|
| 1 | 15 vs 5 | 1 | 0.2159165 | 5.191 | 0.36 | 0.052 | 0.052 | . |
| 2 | 15 vs 0 | 1 | 0.8749702 | 151.681 | 0.93 | 0.001 | 0.003 | ** |
| 3 | 5 vs 0 | 1 | 0.1861974 | 5.028 | 0.358 | 0.026 | 0.039 | * |

***Table S3 : Pairwise PERMANOVA results for group (A) and time (B) factors in Columbia.***

(A)

| Group | Pairs | | Df | SumsOfSqs | F.Model | R2 | p.value | p.adjusted |
|---|---|---|---|---|---|---|---|---|
| 1 | ALer-treated15 vs | ALer-treated5 | 1 | 0.0006823338 | 1.203 | 0.23 | 0.5 | 0.50 |
| 2 | ALer-treated15 vs | ALer-untreated | 1 | 0.0849479441 | 110.727 | 0.96 | 0.1 | 0.10 |
| 3 | ALer-treated15 vs | BLer-treated15 | 1 | 0.0160558036 | 2.868 | 0.41 | 0.1 | 0.107 |
| 4 | ALer-treated15 vs | BLer-treated5 | 1 | 0.6973550840 | 206.366 | 0.98 | 0.1 | 0.10 |
| 5 | ALer-treated15 vs | BLer-untreated | 1 | 0.1167406590 | 45.995 | 0.91 | 0.1 | 0.10 |
| 6 | ALer-treated5 vs | ALer-untreated | 1 | 0.0877770089 | 97.036 | 0.96 | 0.1 | 0.10 |
| 7 | ALer-treated5 vs | BLer-treated15 | 1 | 0.0135712315 | 2.366 | 0.37 | 0.1 | 0.10 |
| 8 | ALer-treated5 vs | BLer-treated5 | 1 | 0.6788136053 | 193.030 | 0.97 | 0.1 | 0.10 |
| 9 | ALer-treated5 vs | BLer-untreated | 1 | 0.1184681461 | 44.278 | 0.91 | 0.1 | 0.10 |
| 10 | ALer-untreated vs | BLer-treated15 | 1 | 0.0582138058 | 9.809 | 0.71 | 0.1 | 0.10 |
| 11 | ALer-untreated vs | BLer-treated5 | 1 | 0.6624979056 | 178.234 | 0.97 | 0.1 | 0.10 |
| 12 | ALer-untreated vs | BLer-untreated | 1 | 0.0136732853 | 4.754 | 0.54 | 0.1 | 0.10 |
| 13 | BLer-treated15 vs | BLer-treated5 | 1 | 0.5667957709 | 66.317 | 0.94 | 0.1 | 0.10 |
| 14 | BLer-treated15 vs | BLer-untreated | 1 | 0.0716811071 | 9.302 | 0.69 | 0.1 | 0.10 |
| 15 | BLer-treated5 vs | BLer-untreated | 1 | 0.5383441707 | 98.096 | 0.96 | 0.1 | 0.10 |

(B)

| Time | Pairs | Df | SumsOfSqs | F.Model | R2 | p.value | p.adjusted | sig |
|---|---|---|---|---|---|---|---|---|
| 1 | 15 vs 5 | 1 | 0.2917675 | 3.989581 | 0.28 | 0.112 | 0.112 | ns |
| 2 | 15 vs 0 | 1 | 0.1509272 | 23.723086 | 0.70 | 0.004 | 0.012 | * |
| 3 | 5 vs 0 | 1 | 0.3573002 | 4.975931 | 0.33 | 0.032 | 0.048 | * |

***Table S4: Pairwise PERMANOVA results for group (A) and time (B) factors in Landsberg.***

| Species type | Amplified region | Primer |
|---|---|---|
| Fungi | ITS1-1F | CTTGGTCATTTAGAGGAAGTAA.GCTGCGTTCTTCATCGATGC |

***Table S5: IST1-1F primer used in this study***

| Sample | RawPE | Combined | Qualified | Nochime | Base(nt) | Avglen(nt) | GC | Q20 | Q30 |
|---|---|---|---|---|---|---|---|---|---|
| AITSColC1 | 105743 | 105111 | 104087 | 103067 | 25580227 | 248.19 | 53.14% | 99.23% | 97.79% |
| AITSColC2 | 109893 | 109368 | 108476 | 107807 | 26934170 | 249.84 | 52.98% | 99.36% | 97.99% |
| AITSColC3 | 104819 | 104235 | 103477 | 102981 | 25693985 | 249.5 | 52.98% | 99.24% | 97.78% |
| AITSCol51 | 105276 | 98934 | 86507 | 85664 | 25197001 | 294.14 | 51.59% | 93.57% | 86.42% |
| AITSCol52 | 56785 | 53341 | 43443 | 43072 | 14785144 | 343.27 | 54.75% | 91.64% | 82.80% |
| AITSCol53 | 106347 | 99999 | 82761 | 81968 | 28331658 | 345.64 | 54.97% | 91.60% | 82.70% |
| AITSCol151 | 104220 | 99195 | 87333 | 86742 | 25811835 | 297.57 | 52.73% | 93.38% | 85.91% |
| AITSCol152 | 105839 | 104183 | 100360 | 100018 | 24934469 | 249.3 | 52.67% | 97.84% | 95.02% |
| AITSCol153 | 111386 | 104151 | 85400 | 84584 | 27372192 | 323.61 | 53.72% | 92.33% | 84.25% |
| AITSLerC1 | 67418 | 67192 | 67081 | 66248 | 16181475 | 244.26 | 53.02% | 99.56% | 98.40% |
| AITSLerC2 | 92592 | 92289 | 92133 | 91241 | 22348582 | 244.94 | 53.01% | 99.59% | 98.45% |
| AITSLerC3 | 105349 | 105032 | 104782 | 103396 | 25303224 | 244.72 | 53.01% | 99.65% | 98.57% |
| AITSLer51 | 102765 | 102359 | 102056 | 101766 | 25283761 | 248.45 | 52.90% | 99.58% | 98.42% |
| AITSLer52 | 112642 | 112141 | 111622 | 111293 | 27692258 | 248.82 | 52.88% | 99.39% | 98.07% |
| AITSLer53 | 104659 | 104162 | 103614 | 103164 | 25574620 | 247.9 | 52.86% | 99.49% | 98.20% |
| AITSLer151 | 86392 | 86014 | 85724 | 82311 | 20450200 | 248.45 | 52.88% | 99.45% | 98.14% |
| AITSLer152 | 113865 | 113426 | 113064 | 112772 | 28012031 | 248.4 | 52.87% | 99.46% | 98.19% |
| AITSLer153 | 103045 | 102658 | 102395 | 100011 | 24794808 | 247.92 | 52.87% | 99.62% | 98.47% |
| BITSColC1 | 105853 | 105464 | 105256 | 103827 | 25394807 | 244.59 | 52.85% | 99.52% | 98.30% |
| BITSColC2 | 106354 | 106009 | 105786 | 104635 | 25605711 | 244.71 | 52.87% | 99.67% | 98.62% |
| BITSColC3 | 102343 | 102019 | 101766 | 100851 | 24841533 | 246.32 | 52.83% | 99.66% | 98.58% |
| BITSCol51 | 102328 | 101391 | 99492 | 99147 | 25312362 | 255.3 | 53.01% | 98.82% | 97.02% |
| BITSCol52 | 103362 | 102133 | 99107 | 98766 | 25576678 | 258.96 | 53.12% | 98.16% | 95.77% |
| BITSCol53 | 104130 | 99537 | 84058 | 83419 | 26070548 | 312.53 | 53.88% | 93.01% | 85.97% |
| BITSCol151 | 109278 | 103468 | 86538 | 85778 | 27422600 | 319.69 | 53.72% | 92.16% | 83.58% |
| BITSCol152 | 103809 | 98266 | 82743 | 82107 | 25310587 | 308.26 | 53.02% | 92.92% | 85.15% |
| BITSCol153 | 100193 | 95904 | 81340 | 79986 | 25753300 | 321.97 | 53.27% | 92.33% | 84.32% |
| BITSLerC1 | 103016 | 102632 | 102398 | 101056 | 24640763 | 243.83 | 52.98% | 99.66% | 98.61% |
| BITSLerC2 | 116539 | 116167 | 116000 | 114157 | 27694508 | 242.6 | 53.04% | 99.72% | 98.77% |
| BITSLerC3 | 102311 | 101962 | 101765 | 100295 | 24453959 | 243.82 | 53.00% | 99.66% | 98.60% |
| BITSLer51 | 86198 | 85734 | 85359 | 84768 | 20384856 | 240.48 | 52.13% | 99.36% | 97.90% |
| BITSLer52 | 112693 | 112043 | 111799 | 111505 | 25968449 | 232.89 | 52.03% | 99.46% | 97.99% |
| BITSLer53 | 106804 | 106025 | 104751 | 104656 | 24848541 | 237.43 | 51.95% | 98.93% | 97.02% |
| BITSLer151 | 155168 | 151611 | 140705 | 139995 | 39388239 | 281.35 | 53.02% | 96.09% | 91.75% |
| BITSLer152 | 102368 | 101022 | 97908 | 97648 | 25239193 | 258.47 | 53.14% | 98.33% | 96.02% |
| BITSLer153 | 106543 | 105060 | 101804 | 101456 | 26300258 | 259.23 | 53.14% | 98.16% | 95.71% |

***Table S6: Summarization of each step of processing for sequencing data of fungal ITS. RawPE (250 bp paired-end raw reads) represents original PE reads after sequencing; Combined represents tags merged from PE reads; Clean represents tags after filtering; Effective represents tags after filtering chimera and can be finally used for subsequent analysis; Base is the number of bases of the Effective Tags; AvgLen represents average length of Effective Tags; Q20 and Q30 are the percentages of bases whose quality value in Effective Tags is greater than 20 (sequencing error rate is less than 1%) and 30 (sequencing error rate is less than 0.1%); GC (%) represents GC content in Effective Tags. Col. Ecotype Columbia; Ler. ecotype Lansberg; C. intreated control; 5. 5 min CAP treatment; 15. 15 min CAP treatment. Replicates are indicated as 1. 2 and 3***

## 6. Author contributions

L.T. performed experiments, analyzed and interpreted data. T.D. and C.B. designed the research. C.K. contributed to design and analyze nephelometric measurements. N.C. performed statistical analyses with L.T. L.T. wrote the paper with contributions of T.D. and C.B. All authors have read and agreed to the published version of the manuscript.

## 7. Funding

This work was supported by a PhD grant from Sorbonne Université (IPV programme) and received financial state aid as part of the PF2ABIOMEDE platform co-funded by «Région Ile-de-France» (Sesame, Ref. 16016309) and Sorbonne Université (technological platforms funding).

## 8. Data Availability Statement

The dataset supporting the conclusions of this article is available in the BioProject database under ID PRJNA1445113 (https://www.ncbi.nlm.nih.gov/sra/PRJNA1445113, accessed on 1 September 2026).

## 9. Acknowledgments

The authors wish to thank Mathieu Barret and Marie Simonin (IRHS Angers, France) for their help and critical comments.

## 10. Conflicts of interest

The authors declare no conflicts of interest.